\documentstyle[12pt]{article}
\begin{document}

\title{Noncommutative Unification of General Relativity and
Quantum Mechanics. A Finite Model.}

\author{Michael Heller\thanks{Correspondence address: ul.
Powsta\'nc\'ow Warszawy 13/94, 33-110 Tarn\'ow, Poland. E-mail:
mheller@wsd.tarnow.pl} \\
Vatican Observatory, V-00120 Vatican City State
\and Zdzis{\l}aw Odrzyg\'o\'zd\'z \and
Leszek Pysiak \and Wies{\l}aw Sasin \\
Department of Mathematics and Information Science, \\ Warsaw
University of 
Technology \\
Plac Politechniki 1, 00-661 Warsaw, Poland}
\date{\today}
\maketitle

\begin{abstract}
We construct a model unifying general relativity and quantum
mechanics in a broader structure of noncommutative geometry. The
geometry in question is that of a transformation groupoid
$\Gamma $ given by the action of a finite group on a space $E$.
We define the algebra ${\cal A}$ of smooth complex valued
functions on $\Gamma $, with convolution as multiplication, in
terms of which the groupoid geometry is developed. Owing to the
fact that the group $G$ is finite the model can be computed in
full details. We show that by suitable averaging of
noncommutative geometric quantities one recovers the standard
space-time geometry. The quantum sector of the model is explored
in terms of the regular representation of the algebra ${\cal
A}$, and its correspondence with the standard quantum mechanics
is established.
\end{abstract}

\section{Introduction}
There are many attempts to create a quantum gravity theory, or
at least some kind of unification of general relativity and
quantum mechanics, based on noncommutative geometry (see, for
example, \cite{Chamsedetal,Sitarz,Connes96,MadMour,MadSae}). In
a series of works \cite{HSL,Emergence,Towards,Reduction} we have
also proposed a model aimed at a unification of relativity and
quanta which differs from other models of this type by the ample
use of the groupoid concept. We consider a transformation
groupoid $\Gamma = E \times G$, where $E$ is typically the frame
bundle over space-time $M$ and $G$ a group acting on $E$, and
the noncommutative algebra ${\cal A}$ of compactly supported
complex valued functions on $\Gamma $ with convolution as
multiplication. As a next step we develop the geometry of the
groupoid based on the algebra ${\cal A}$ and its derivations.
The $E$-component of this geometry reconstructs standard general
relativity, and its $G$-component is interpreted as a quantum
sector of the model. The natural choice for $G$ is the Lorentz
group or some of its representations.
\par
Preliminary results indicate that the model is worth exploring
as a possible step in the right direction. However, it is
involved in many mathematical intricacies that often overshadow
interconceptual relations, and at this stage exactly these
relations are especially important. It turns out that if $G$ is
assumed to be a finite group, the model becomes ``fully
computable'' and all conceptual issues are clarified. To
construct such a model is the aim of the present work. Because
of the finiteness of $G$ we call this model a finite model,
although other its ``components'' remain infinite. In
particular, $M$ can be any relativistic space-time.
\par
We compute such a model in all its details. Some of our previous
results have been confirmed, and some new have emerged.  The
most interesting aspects of the model concern the architecture
of the groupoid geometry, the structure of Einstein equations,
and dynamics in the quantum sector. The nice result is that the
transition from the noncommutative geometry of our model to the
classical space-time geometry can be done by averaging elements
of the algebra ${\cal A}$ in analogy to what is done in the
usual quantum mechanics.
\par
We begin our analysis with a brief reminder, in Section 2, of
the transformation groupoid structure (mainly to fix notation).
In Section 3, we discuss the noncommutative algebra ${\cal A}$
on the transformation groupoid $\Gamma = E \times G$ where $G$
is a finite group and, in Section 4, we develop the geometry of
the groupoid $\Gamma $ based on the algebra ${\cal A}$ and the
module of its derivations. As a simple but instructive example
we compute, in Section 5, the geometry of the groupoid with $G =
{\bf Z}_2$. Section 6 is devoted to establishing the
correspondence between the geometry of our model and the
classical space-time geometry via the above mentioned averaging
procedure. In Section 7, we explore the quantum sector of our
model in terms of the regular representation of the algebra
${\cal A}$ and discuss its correspondence with quantum
mechanics. Main results are collected in Section 8.

\section{A Transformation Groupoid}
In this section we give a brief description of the groupoid
structure mainly to fix notation (for details see, for instance,
\cite[chapter 1]{Paterson}). {\it Groupoid\/} is a set $\Gamma
$ with a distinguished subset $\Gamma ^2 \subset \Gamma
\times \Gamma $, called the {\it set of composable elements\/},
equipped with two mappings:

$\cdot :\Gamma^2 \rightarrow \Gamma $ defined by $(x, y)
\mapsto x \cdot y, $ called {\it multiplication\/}, and
 
$^{-1}:\Gamma \rightarrow \Gamma $ defined by $x \mapsto x^{-1}$
such that $(x^{-1})^{-1} = x$, called {\it inversion\/}.

These mappings have the following properties

(i) if $(x,y),(y,z) \in \Gamma^2$ then $(xy, z), (x, yz) \in
\Gamma^2 $ and $(xy)z = x(yz)$,

(ii) $(y,y^{-1}) \in \Gamma^2 $ for all $y \in \Gamma $, and if
$(x,y) \in \Gamma^2$ then $ x^{-1}(xy) = y$ and $(xy)y^{-1} =
x.$

One also defines the {\it set of units\/} $\Gamma^0 =
\{x^{-1}x: x \in \Gamma \} \subset \Gamma $, and the two
following mappings:  $d, r: \Gamma \rightarrow \Gamma^0 $ by
$d(x) = x^{-1}x,$ and $r(x) = xx^{-1}$, called the {\it source
mapping\/} and the {\it target mapping}, respectively.  Two
elements $x$ and $y$ can be multiplied, i.e.,  $(x,y) \in
\Gamma^2$, if and only if $d(x) = r(y)$.  
For each $u \in \Gamma^0$ one defines the sets $$\Gamma_u = \{x
\in \Gamma: d(x) = u\} = d^{-1}(u) $$ and $$\Gamma^u = \{x \in
\Gamma: r(x) = u\} = r^{-1}(u).$$ Both these sets give different
fibrations of $\Gamma $.

The above purely algebraic construction can be equipped with the
smoothness structure. If this is the case, it is called a {\it
smooth\/} or {\it Lie groupoid\/} \cite[chapter 2.3]{Paterson}.

Let $\tilde{E}$ be a differential manifold (or a differential
space, see \cite{Structured}) with a group $\tilde{G}$ acting on
it smoothly and freely to the right, $\tilde E \times \tilde G
\rightarrow \tilde E$. This action leads to the bundle
$(\tilde{E}, \pi_M , M = \tilde{E}/\tilde{G})$. The special case
of this construction is the frame bundle over $M$ with the
Lorentz group $\tilde{G}$ as its structural group. Let now $G$
be a finite subgroup of $\tilde{G}$, and let $S:M \rightarrow
\tilde E$ be a cross section of the bundle $(\tilde{E}, \pi_M ,
M)$. We do not assume that this cross section must be continuous
(we simply chose one element of $E$ from each fibre).  Now, we
define $E = \bigcup_{x\in M}S(x)G$. We understand $E$ as a
differential space $(E, C^{\infty }(\tilde E)_E)$.

$G$ acts freely (to the right) on $E$, $E\times G \rightarrow
E$, which gives the groupoid structure to the Cartesian product
$
\Gamma = E \times G$.  It is a special case of {\em
transformation groupoids\/} and will constitute the subject
matter of the present study.  The elements of $\Gamma $ are
pairs $\gamma =(p,g)$ where $p\in E$ and $ g\in G$. Two such
pairs $\gamma_1 = (p, g)$ and $\gamma_2 =(pg, h)$ are composed
in the following way $$
\gamma_2 \circ \gamma_1 = (pg, h)(p, g) = (p, gh).
$$ The inverse of $(p,g)$ is $(pg, g^{-1}).$ The set of units is
$$
\Gamma^0 = \{\gamma^{-1} \gamma: \gamma \in \Gamma \} = \{(p,
e):\, p \in E \}.  $$

We could think of $\gamma =(p,g)$ as of an arrow beginning at $
p$ and ending at $pg$. Two arrows $\gamma_1$ and $\gamma_2$ can
be composed if and only if the beginning of $\gamma_2$ coincides
with the end of $\gamma_1.$

Let us notice that if the cross section $S: M \rightarrow \tilde
E$ is smooth, the bundle $(\tilde E, \pi_M, M)$, where $\pi_M$
is the canonical projection $\pi_M: \tilde E\rightarrow M$, is a
trivial $\tilde G$-bundle.  Indeed, the trivializing
diffeomorphism $\phi: \tilde E
\rightarrow M\times \tilde G$ is given by $\phi (p) = (\pi_M(p),
g_p)$ where $g_p$ is the element of the group $\tilde G$ such
that $p=S(\pi_M(p)) g_p$.

Let us also notice that the source and range mappings for
$\gamma = (p,g)$ can now be written as
\[
d(\gamma ) = p = S(x) \cdot g_1, \] \[ r(\gamma ) = pg =
S(x)\cdot g_2,
\] $x\in M$, for $g_1, g_2 \in G$, respectively; of course, $g_2
= g_1 g$.

\section{The Groupoid Algebra}
We define the algebra ${\cal A}= C^{\infty }(\Gamma , {\bf C})$
of smooth complex valued functions on the groupoid $\Gamma = E
\times G$ with the co\-nv\-olution as mul\-ti\-pli\-cation. If $f, g \in
{\cal A}$, the con\-vo\-lution is defined as
\[
(f*g)(\gamma ) = \sum_{\gamma_1 \in \Gamma_{d(\gamma )}}f(\gamma
\circ \gamma_1^{-1})g(\gamma_1).
\]
\par
Let us define the mapping $\varphi : \Gamma \rightarrow
\bigcup_{x \in M} E_x \times E_x, $ where $E_x = \pi_M^{-1}(x)$,
by $\varphi (\gamma )= (p_0, p_1)$ with $p_0 = d(\gamma )$ and
$p_1 = r(\gamma )$. Here we identify $E$ with the set $\Gamma^0$
of units of the groupoid $\Gamma $. It can be easily seen that
$\varphi $ is a bijection. If we introduce the abbreviation $
\tilde f(p_0, p_1) = f(\varphi^{-1}(p_0, p_1)), $ the
convolution is expressed by
\begin{eqnarray*}
(\tilde f * \tilde g)(p_0, p_1) & = & \sum_{p \in E_x}\tilde
f[(p_0,p_1)\circ (p, p_0)]\tilde g(p_0, p) \\ & = & \sum_{p\in
E_x}\tilde f(p,p_1) \tilde g(p_0,p).
\end{eqnarray*}
Here $x = \pi_M(p_0)$.

Now, with a function
\[
\phi : M \times G \times G \rightarrow {\bf C} \]
we associate the matrix $A_{\phi }$ given by
\[
A_{\phi }(\cdot ,i,j) = \phi (\cdot , g_i, g_j).
\]
The function $\phi $ allows us to define the mapping
\[
A_{\phi }: M \times \{1,2, \ldots , k\} \times
\{1,2, \ldots , k\} \rightarrow {\bf C}
\]
with the help of the formula
\[
A_{\phi }(x,i,j) = \phi (x, \sigma (i), \sigma (j))
\]
where $\sigma $ is a bijection given by $\sigma(i) = g_i$, and
$k = |G|$.

It is easy to see that $M \times G \times G$ is a groupoid with
the multiplication
\[
(x,g', \bar{g})\circ (x, g, g') = (x, g, \bar{g}).\]

{\it Lemma\/}. The mappings
\[
\varphi : \Gamma \rightarrow \bigcup_{x \in M} E_x \times E_x
\]
defined above, and
\[
\Phi : \Gamma \rightarrow M\times G
\times G
\] 
given by
\[
\Phi (\gamma) = (pr(\gamma ), \lambda (d(\gamma )), \lambda
(r(\gamma )))
\]
are isomorphisms of groupoids. Here $pr: \Gamma \rightarrow M$
is the natural projection and  $\lambda :E \rightarrow G$ is a
mapping given by $ \lambda (p) = g$ such that $p = S(x)g$.

{\it Proof.\/} Let us prove this for $\Phi $. The mapping
$\Phi^{-1}(x, g_1, g_2) = (S(x)g_1, g_1^{-1}g_2)$ determines the
bijection between $\Gamma $ and $M
\times G \times G $. $\Phi $ is also a homomorphism. Indeed,
\begin{eqnarray*}
\Phi (\gamma_1 \circ \gamma_2 ) & = & (pr(\gamma_1 \circ
\gamma_2), 
\lambda (d(\gamma_1 \circ \gamma_2)), \lambda (r(\gamma_1
\circ \gamma_2)) \\ & = &
(x, g', \bar{g}) \circ (x, g, g') =
\Phi(\gamma_1)\circ \Phi (\gamma_2)
\end{eqnarray*}
where we have introduced the following abbreviations: $\bar{g}=
\lambda (r(\gamma_1)),\/ g = \lambda (d(\gamma_2)), \, g' =
\lambda (d(\gamma_1))$.

By using the mapping $\Phi $ one readily shows that the set of
elements composable in $\Gamma $ is bijective with the set of
elements composable in $M\times G \times G$. It remains to check
the invertibility
\begin{eqnarray*}
\Phi (\gamma^{-1}) & = & (x, \lambda(r(\gamma )), \lambda (d(\gamma
))) \\ & = & (x, \lambda (d(\gamma )), \lambda (r(\gamma
))^{-1}\\ & = & [\Phi(\gamma)]^{-1}.
\end{eqnarray*}
The proof for $\varphi $ is analogous.  $\Box $

{\em Lemma.} The mapping $\Phi^{*}:C^{\infty}(M\times G\times
G)\rightarrow C^{\infty}(\Gamma )$ is an isomorphism of
algebras.

{\em Proof.\/} Since $\Phi^{*}$ is a bijection it is enough to
show that it is a homomorphism
\begin{eqnarray*}
(\phi *\psi )(x,g,\bar {g}) & = & \sum_{g'\in G}\phi (x,g',
\bar {g})\cdot\psi (x,g,g') \\
& = & \sum_{g'\in G}\phi [(x,g.\bar {g})\circ (x,g',g)]\psi
(x,g,g') \\ & = & \sum_{g'\in G}\phi [(x,g,\bar {g})\circ
(x,g,g')^{-1} ]\psi (x,g,g') \\ & = &
\sum_{\gamma_1\in\Gamma_{d(\gamma )}}\phi [\Phi (\gamma
\circ\gamma^{-1}_1)]\psi (\Phi (\gamma_1)) \\
 & = & \sum_{\gamma_1\in\Gamma_{d(\gamma )}}(\Phi^{*}\phi )
(\gamma\circ\gamma_1^{-1})(\Phi^{*}\psi )(\gamma_1) \\ & = &
\sum_{\gamma_1\in\Gamma_{d(\gamma )}}a(\gamma\circ\gamma_
1^{-1})b(\gamma_1).
\end{eqnarray*}
In the last line the obvious abbreviations are introduced. $
\Box$
\par
We can interpret the algebra ${\cal A}$ as the matrix algebra by
defining the following mapping
\[
a \mapsto A_a = [a(x, g_i, g_j)]^k_{i,j = 1}.
\]
The indices $i$ and $j$ label rows and columns of the respective
matrix. In this representation the convolution becomes the usual
matrix multiplication
\[
A_{f*g} = A_g \cdot A_f.
\]
\par
If we remember that the center ${\cal Z}({\cal A})$ of the
algebra ${\cal A}$ is
\[{\cal Z}({\cal A})=\{\overline {\pi_M^{*}f}: f
\in C^{\infty}(M)\},\]
where
\[\overline {\pi_M^{*}f}=\left\{\begin{array}{ll}
0&\mbox{if $\gamma\in E_x\times \{g\},\,x \in M,\,g\neq e$}\\
f(x)&\mbox{if $\gamma\in E_x\times \{g\},\,x\in M,\,
g=e$}\end{array}
\right.,\]
we have the isomorphism of algebras $\zeta : C^{\infty}(M)
\rightarrow {\cal 
Z}({\cal A})$ given by
\[
\zeta(f\cdot {\bf I}) = \overline{\pi_M^*(f)}.
\]

\section{Geometry of the Groupoid}
Let us consider the ${\cal Z}({\cal A})$-module of derivations
of the algebra ${\cal A}$
\[V\equiv {\rm D}{\rm e}{\rm r}{\cal A}={\rm O}{\rm u}{\rm t}
{\cal A}\oplus {\rm I}{\rm n}{\rm n}{\cal A}\] where
\[V_1\equiv {\rm O}{\rm u}{\rm t}{\cal A}:=\{\bar {X}\in V:\bar
{X}(a)=\Phi ^*(X(\Phi^*)^{-1}(a)),\forall X\in {\cal X}(M)\},\]
\[V_2\equiv {\rm I}{\rm n}{\rm n}{\cal A}:=\{{\rm a}{\rm d}
a:a\in {\cal A}\}\] and ad$a(b)=[a,b]$ for $b\in {\cal A}$. We
have
\[[\bar {X},{\rm a}{\rm d}a]={\rm ad}\bar X(a),\;[\bar {X},\bar
{Y}]=\overline { [X,Y]},\;[{\rm a}{\rm d}a,{\rm a}{\rm d}b]={\rm
a}{\rm d} [a,b].\]
\par
This allows us to define the following metric on V
\[{\cal G}(u,v)=\bar {g}(u_1,v_1)+h(u_2,v_2)\]
where $\bar {g}:V_1\times V_1\rightarrow {\cal Z}({\cal A} )$ is
a ``lifting'' of the metric $g:{\cal X}(M)\times {\cal X}
(M)\rightarrow C^{\infty}(M)$ on $M$
\[\bar {g}(\bar {X},\bar {Y})=\overline {g(X,Y)}=\zeta
(g(X,Y)),\] and $h:V_2\times V_2\rightarrow {\cal Z}({\cal A})$
is a metric on the ``noncommutative part'' of the model.
\par
We have the dual module $V^{*}={\rm H}{\rm o}{\rm m}_{{\cal Z}
({\cal A})}(V,{\bf C}))$, and since $V$ is a locally free ${\cal
Z}({\cal A})$-module, there is the isomorphism $
\Phi_{{\cal G}}:V\rightarrow V^{*}$ given by
\[\Phi_{{\cal G}}(u)(v)={\cal G}(u,v)=\Phi_{\bar {{\rm g}}}
(u_1)(v_1)+\Phi_h(u_2)(v_2).\]

Now, we can define the {\em preconnection} $\nabla^{*}: V\times
V\rightarrow V^{*}$ with the help of the Koszul formula
\begin{eqnarray*}
(\nabla^{*}_uv)(x) & = & \frac 12[u({\cal G}(v,x))+v({\cal G}
(u,x))-x({\cal G}(u,v)) \\ & & +{\cal G}(x,[u,v])+{\cal
G}(v,[x,u])-{\cal G}(u,[v,x] ),
\end{eqnarray*}
and then the {\em Levi-Civita connection\/} by
\[\nabla =\Phi_{{\cal G}}^{-1}\circ\nabla^{*}.\]
\par
Now, let us introduce the basis ($\bar{\partial}_{\mu} ,e_i)$,
$\mu =0,1,\ldots ,m,\,i=1,\ldots ,n$, in the ${\cal Z}({\cal
A})$-module $V=V_1\oplus V_2$. We have
\[[\bar{\partial}_{\mu},\bar{\partial}_{\nu}]=0,\;[e_i,
e_j]=c_{ij}^ke_k,\;[\bar{\partial}_{\mu},e_i]=0\] with
$c_{ij}^k\in {\bf C}$ (indeed, if we put $ e_i=\mbox{ad}E_i$, we
have $[e_i,e_j]=\mbox{ad}[E_i,E_j]={\rm a}{\rm d}(c_{ij}^kE_
k)$).
\par
Connection $\nabla$ determines the curvature tensor
\[R(u,v)w=\nabla_u\nabla_vw-\nabla_v\nabla_uw-\nabla_{[
u,v]}w.\] Let us notice that $R(u,v)w=0$ if $u,v,w\in
\{\bar{\partial}_ 0,\ldots ,\bar{\partial}_m,e_1,\ldots ,e_n\}$
and $u,v,w$ do not belong simultaneously to the sets
$\{\bar{\partial}_0 ,\ldots ,\bar{\partial}_m\}$ or
$\{{e_1,\ldots ,e_n}\}$. Consequently, we have
\begin{eqnarray*}
R(u,v)w & = & R(u_1+u_2,v_1+v_2)(w_1+w_2) \\ & = &
R((u^{\alpha}\bar{\partial}_{\alpha}+u^ie_i),(v^{\beta}
\bar{\partial}_{\beta}+v^je_j))(w^{\gamma}\bar{\partial}_{
\gamma}+w^ke_k) \\
& = &
R_{\alpha\beta\gamma}^{\mu}u^{\alpha}v^{\beta}w^{\gamma}\bar{\partial
}_{\mu } + R_{ijk}^ku^iv^jw^ke_l
\end{eqnarray*}
where $R_{\alpha\beta\gamma}^{\mu}\in C^{\infty}(M)$ are the
components of the curvature tensor $\stackrel {
\bar {g}}R$ of the 
connection $\stackrel {\bar {g}}{\nabla}$ and $R_{ijk}^ l$ are
the components of the curvature tensor $\stackrel hR$ of the
connection $\stackrel h{\nabla}$. Therefore,
\[R(u,v)w=\stackrel {\bar {g}}R(u_1,v_1)w_1+\stackrel h
R(u_2,v_2)w_2.\]
\par
This decomposition is, in general, valid for other geometric
magnitudes. We should notice that the moduli ${\cal X}( M)$ and
${\rm O}{\rm u}{\rm t}{\cal A}$ are isomorphic which means that
the geometry of $\bar {g}$ is a copy of that of $ g$ on $M$.
Therefore, in the $\bar g$-sector the situation is the same as
in the usual general relativity: we can formulate Einstein
equations that are to be solved for the metric. In the
$h$-sector of our model the situation is different; let us
analyze it in the more detailed way.
\par
Having a basis the trace of a ${\cal Z}({\cal A})$-endomorphism
$ A:V_2\rightarrow V_2$ is defined in the usual way: ${\rm
t}{\rm r}A=\sum^n_{i=1}A^i_i\in {\cal Z} ({\cal A})$. For a
fixed pair $x,y\in V_2$ one defines the family of operators
$\stackrel hR_{xy}:V_2\rightarrow V_ 2$ by
\[\stackrel hR_{xy}(v)=\stackrel hR(v,x)y,\]
and the Ricci 2-form ${\bf r}{\bf i}{\bf c}_h:V_2\times
V_2\rightarrow {\cal Z}({\cal A})$ by
\[{\bf r}{\bf i}{\bf c}_h(x,y)={\rm t}{\rm r}\stackrel 
hR_{xy},\] or in the local basis
\[{\bf r}{\bf i}{\bf c}_h(u,v)=\stackrel hR_{ij}u^iv^j\] where
$u=u^ie_i,\,v=v^je_j$. There exists uniquely defined operator
$\stackrel h{{\cal R}}$: $V_2\rightarrow V_2$ given by
\[{\bf r}{\bf i}{\bf c}_h(u,v)=h(\stackrel h{\cal R}(u),v).\]
And the scalar curvature is defined as
\[\stackrel hr={\rm t}{\rm r}\stackrel h{\cal R}.\]
\par
Now, we have all quantities required to write the counterpart of
the usual Einstein equation
\[\stackrel h{{\cal R}}-\frac 12\stackrel hr{\rm i}{\rm d}_{
V_2}+\Lambda {\rm i}{\rm d}_{V_2}=\kappa T.\] Here $\Lambda$ and
$\kappa$ are counterparts of the cosmological constant and
Einstein's gravitational constant, respectively, and $T$ is a
counterpart of the energy-momentum tensor. Since, however, our
philosophy is that matter should be generated out of ``purely
noncommutative geometry'', we prefer to consider the above
equation with $T=0$, i.e.,
\[\stackrel h{{\cal R}}+\Lambda {\rm i}{\rm d}_{V_2}=0,\]
or, if we write it with the argument and omit the cumbersome
superscript $h$,
\begin{equation}{\bf G}_h\equiv {\cal R}(u)+\Lambda u=0
.\label{Einh}\end{equation}
Now, it is clear that (\ref{Einh}) is an eigenvalue equation,
and it should be solved with respect to $u\in V_2$. If we assume
that $u=u^ie_i ,\,i=1,\ldots ,n$, with $u^i\in {\cal Z}({\cal
A})$, this equation takes the form
\[u^i{\cal R}_i^j+\Lambda u^j=0.\]
It has nontrivial solutions if
\[{\rm d}{\rm e}{\rm t}({\cal R}_i^j+\Lambda {\bf I})=0
.\]
This implies that $\Lambda\in {\cal Z}({\cal A})$ which means
that $
\Lambda$ is a function on $M$ (it is 
constant only at $x\in M$).
\par
Let us now consider the full Einstein equation on the groupoid
\begin{equation}
{\bf G}_{\bar {g}}+{\bf G}_h=0\label{Ein}
\end{equation}
where ${\bf G}_{\bar {g}}=0$ are the usual Einstein equations on
space-time $ M$ suitably lifted to the groupoid.  It is,
therefore, evident that $
\bar {g}$ solves ${\bf G}_{\bar {g}}=0$ if and 
only if the corresponding metric $g$ solves the usual Einstein
equations on $M$.  Let us notice that the generalized Einstein
equation (\ref{Ein}) determines the pair $(V,{\cal G})$, i.e.,
the module of derivations and the metric on it.  In the case of
the standard geometry on space-time $ M$, the module of
derivations is unique and we are looking for the metric.  This
is also true for equation ${\bf G}_{\bar {g}}=0$, but for
equation (\ref{Einh}) it could be that $h$ is unique (see
\cite[p. 75]{Madore}, and in this case we should solve this
equation for derivations.

\section{A Simple Example} 
In this section we test our approach by considering a simple
example in which $G={\bf Z}_2$, where ${\bf Z}_2 =\{1,\epsilon
\},\,\epsilon^2=1${\bf ,} and $E=M\times {\bf Z}_2$.  Therefore,
we have the groupoid $\Gamma =E\times G${\bf .}  Its elements
are:

$\gamma_1=\gamma_{1,x}=((x,1),1)\stackrel {\Phi}{\rightarrow}
(x,1,1)\in M\times G\times G,$

$\gamma_2=\gamma_{2,x}=((x,1),\epsilon )\stackrel {\Phi}{
\rightarrow}(x,1,\epsilon )\in M\times G\times G,$

$\gamma_3=\gamma_{3,x}=((x,\epsilon ),\epsilon )\stackrel {
\Phi}{\rightarrow}(x,\epsilon ,1)\in M\times G\times G,$

$\gamma_4=\gamma_{4,x}=((x,\epsilon ),1)\stackrel {\Phi}{
\rightarrow}(x,\epsilon ,\epsilon )\in M\times G\times 
G.$

We remember that $\Phi$ is an isomorphism of groupoids. In fact,
we have here a family of groupoids (a groupoid over each $x\in
M$) which is also a groupoid.

Let us now consider the algebra ${\cal A}=(C^{\infty}(\Gamma
,{\bf C}),*)$. If $f\in {\cal A}$, we have
$f_{11}=f_{11,x}=f(\gamma_{1,x})$, and similarly for other
elements. There is the correspondence
\[{\cal A}\ni f\rightarrow M_f=\left[\begin{array}{cc}
f_{11}&f_{12}\\ f_{21}&f_{22}\end{array}
\right]\in C^{\infty}(M)\otimes M_{2\times 2}({\bf C}).\]
For fixed $x\in M$ it is a matrix with numerical entries.
\par
The convolution is antiisomorphism. We have
\begin{eqnarray*}
(f*g)(\gamma_1) & = & (f*g)_{11}=f(\gamma_1\circ\gamma_1^{-
1})g(\gamma_1)+f(\gamma_1\circ\gamma_2^{-1})g(\gamma_2) \\ & = &
f_{11}\cdot g_{11}+f_{21}\cdot g_{12}
\end{eqnarray*} 
which is the matrix multiplication rule. And similarly for other
matrix elements.
\par
The ${\cal Z}({\cal A})$-module of inner derivations $ V_2$ is
isomorphic with $sl_2({\bf C})\otimes C^{\infty} (M)$.  Let us
choose the basis in $sl_2({\bf C})$
\[H_0=\frac 12\left[\begin{array}{cc}
1&0\\ 0&-1\end{array}
\right],\;X_1=\left[\begin{array}{cc}
0&1\\ 0&0\end{array}
\right],\;X_2=\left[\begin{array}{cc}
0&0\\ 1&0\end{array}
\right]\]
which leads to the following commutation relations
\[[H_0,X_1]=X_1,\;[H_0,X_2]=-X_2,\;[X_1,X_2]=2H_0.\]
\par
The natural choice for the metric is the Killing form
\[h(X,Y)=\langle X,Y\rangle =\mbox{Tr$ $}(\mbox{ad}X\circ\mbox{ad}
Y).\] We can easily compute that the only nonvanishing
components of this metric are
\[\langle H_0,H_0\rangle =2\;\mbox{and}\;\langle X_1,X_
2\rangle =4.\] Now, it is convenient to change to the new basis
\[H=\frac 1{\sqrt {2}}H_0,\;Y_1=\frac 1{2\sqrt {2}}(X_1
+X_2),\;Y_2=\frac 1{2\sqrt {2}}(X_1-X_2)\] in which
\[\langle H,H\rangle =\langle Y_1,Y_1\rangle =1,\;\langle 
Y_2,Y_2\rangle =-1\] and
\[\langle H,Y_1\rangle =\langle H,Y_2\rangle =\langle Y_
1,Y_2\rangle =0.\]
\par
Let us introduce the following notation: $\partial_x=\mbox{ad}
x$ where $x$ is a traceless matrix, and
$\partial_i=\mbox{ad}E_i$. It can be shown that the connection
\[\nabla_{\partial_x}\partial_y=\alpha [\partial_x,\partial_
y],\] for any $\alpha\in {\cal Z}({\cal A})$, is compatible with
the Killing metric, i.e.
\[\nabla_{\partial_z}\langle\partial_x,\partial_y\rangle 
=\langle\nabla_{\partial_z}\partial_x,\partial_y\rangle
+\langle\partial_x,\nabla_{\partial_z}\partial_y\rangle
.\]
\par
We now readily compute the curvature tensor
\[R(\partial_x,\partial_y)\partial_z=(\alpha^2-\alpha )
[[\partial_x,\partial_y],\partial_z],\] and the torsion tensor
\[T(\partial_x,\partial_y)=(2\alpha -1)[\partial_{x,}\partial_
y].\] To have $T=0$ we must assume $\alpha =1/2$. Hence, for the
Ricci tensor we have
\[\mbox{{\bf ric}}(\partial_k,\partial_l)=\frac 14\sum_{
j=1}^{N^2-1}\langle [\partial_k,\partial_j],[\partial_l
,\partial_j]\rangle .\] We easily compute that the only
nonvanishing component of the Ricci tensor, in the basis
$(H,Y_1,Y_2)$, is
\[\mbox{{\bf ric}}(Y_2,Y_2)=\frac 14.\]
\par
Finally, the Einstein equation assumes the form of the
eigenvalue equation
\[{\cal R}(u)+\Lambda (u)=0.\]
For the eigenvalue $\Lambda =-1/4$, the space of its solutions
is
\[W=\{u\in V_2:u=f\cdot Y_2,f\in C^{\infty}(M)\}.\]
Of course, $W$ is a ${\cal Z}({\cal A})-$submodule of the module
$ V_2$.  The modular dimension of $W$ is one.  If $\Lambda =0$,
the space of solutions is spanned by $ H$ and $Y_1$, and is of
modular dimension two.

\section{Correspondence with Classical Theory} 
It is interesting to notice that the transition from the
noncommutative geometry on the groupoid $\Gamma$ to the
classical geometry on the manifold $M$ can be done with the help
of an averaging procedure where the averaging of a functional
matrix $A$ is given by
\[
\langle A \rangle = \frac{1}{|G|}{\rm Tr}A.
\]
This averaging kills noncommutativity; indeed
\[ 
\langle AB \rangle = \frac{1}{|G|}{\rm Tr}(AB) =
\frac{1}{|G|}{\rm Tr}(BA) = \langle BA \rangle .
\]
\par
Let $f\in C^{\infty}(M)$ be a function on $M$; it can be
expressed as $A_x=f(x)\cdot {\bf I}$, $x\in M$, and its
averaging gives $\langle A_x\rangle = \frac{1}{|G|}{\rm Tr}(A) =
f(x)$.  In this way, we have demonstrated that functions $f(x)$
on $M$, interpreted as $A_x=f(x)\cdot {\bf I}$, have the
property that the average of $A_x$ is equal to $f(x)$.
\par
Moreover, there exists the mapping ${\rm tr}: {\cal A}
\rightarrow C^{\infty }(M)$ given by
\[{\rm t}{\rm r}a:={\rm T}{\rm r}((\Phi^{-1})^{*}a)\]
for every $a\in {\cal A}$. Indeed, we have $(\Phi^{-1})^*a \in
C^{\infty }(M \times G \times G)$, i.e., $(\Phi^{-1})^*a =
\varphi (x, g_1, g_2)$, and its trace ${\rm Tr}: C^{\infty }(M
\times G \times G) \rightarrow {\bf C}$ is given by
 \[ ({\rm Tr}\varphi )(x) = \sum_{g \in G}\varphi(x, g, g).\] It
is easy to check that, besides the usual properties of trace,
one has
\[ {\rm tr}(\varphi * \psi) = {\rm tr} (\psi * \varphi ) \] 
for $\varphi , \psi \in C^{\infty }(M \times G \times G)$. Let
us also notice that
\[ \frac{1}{|G|}{\rm tr}|{\cal Z}({\cal A}) = \zeta^{-1}.\]
\par

{\it Lemma.\/} There exists the canonical projection $P: {\cal
A} \rightarrow {\cal Z}({\cal A})$, $P = P^2$ and $P$ is ${\bf
C}$-linear, such that $P|{\cal Z}({\cal A}) = {\rm id}_{{\cal
Z}({\cal A})}$.

{\it Proof.\/} We define $P: {\cal A} \rightarrow {\cal Z}({\cal
A})$ by
\[
P = \zeta \circ \frac{1}{|G|}{\rm tr},
\]
and easily check its properties formulated in the Lemma.  $\Box
$.
\par
It can be easily seen that for any $ u\in {\rm D}{\rm e}{\rm
r}{\cal A}$ and any element $a\in ({\cal Z}({\cal A}))$ we have
$u(a) \in {\cal Z}({\cal A})$. For every $u
\in {\cal Z}({\cal A})$ we define the projection $u^\#:
C^{\infty }(M) \rightarrow C^{\infty }(M)$ by
\[ u^\#(f) = \zeta^{-1}(u(\zeta(f))). \]
If $u$ is an inner derivation, then $({\rm ad}b)^\# = 0$ for any
$b \in {\cal A}$, and for any $X \in {\cal X}(M))$ one has
${\bar{X}}^\#=X$.
\par
We see that ${\rm D}{\rm e}{\rm r}{\cal A}\ni u \mapsto u^\#
\in {\rm D}{\rm e}{\rm r}(C^{\infty}(M))$ is a homomorphism of
Lie algebras, and its restriction to the center $u|{\cal
Z}({\cal A})\mapsto u^\#$ is an isomorphism of Lie algebras.

Let $\omega$ be a $k$-form
\[\omega : \underbrace {{\rm D}{\rm e}{\rm r}{\cal A}\times
\cdots\times {\rm D}{\rm e}{\rm r}{\cal A}}_{k\;times}
\rightarrow {\cal Z}({\cal A}),\]
then $\omega ^\# ={\cal X}(M)\times\cdots\times {\cal
X}(M)\rightarrow C^{\infty}(M)$ is given by
\[\omega ^\# (X_1,\ldots ,X_k)=\zeta^{-1}(\omega 
(\bar {X}_1,\ldots ,\bar {X}_k))=(\omega (\bar {X}_1,\ldots\bar
{X}_k))^\#.\]
\par
Similarly, for the connection $\nabla : {\rm D}{\rm e} {\rm
r}{\cal A}\times {\rm Der}{\cal A}\rightarrow {\rm Der}{\cal A}$
we have $\nabla ^\#:{\rm D}{\rm e}{\rm r}(C^{\infty}
(M))\times\,{\rm D}{\rm e}{\rm r}(C^{\infty}(M))\rightarrow {\rm
D}{\rm e}{\rm r}(C^{\infty}(M))$ given by
\[\nabla ^\#_XY= (\nabla_{\bar {X}}\bar {Y})^\#.\]
\par 
And generally for a tensor $A:{\rm D}{\rm e}{\rm r}{\cal A}
\times\cdots \times {\rm D}{\rm e}{\rm r}{\cal A}\rightarrow
{\rm D} {\rm e}{\rm r}{\cal A}$ we obtain $A^\#: {\rm D}{\rm
e}{\rm r}(C^{\infty}(M))
\times\cdots\times\,{\rm D}{\rm e}{\rm r}(C^{\infty}(M)
)\rightarrow {\rm D}{\rm e}{\rm r}(C^{\infty}(M))$ which is
given by
\[A^\#(X_1,\ldots ,X_k)= (A(\bar {X}_1,\ldots ,\bar
{X}_k))^\#.\]
\par
Consequently, for our metric we have
\[(\bar {g}+h)^\#(X_1,X_2)= (\bar {g}(\bar {
X}_1,\bar {X}_2))^\#=\overline{ ({ g(X_1,X_2))^\#}}
=g(X_1,X_2)\] for all $X_1,X_2\in {\cal X}(M)$, as it should be.
This is obvious if one remembers that $h(\bar {X}_1,\bar
{X}_2)=0$.

Therefore, we can say that the usual differential geometry on
the base manifold $M$ is the averaging of the differential
geometry developed in Section 4. This averaging corresponds to
the averaging with respect to units of the groupoid (which, in
the matrix representation of the groupoid, is equivalent to the
averaging of the diagonal elements of a given matrix).

\section{Regular Representation and Quan\-tum Sec\-tor of the
Model} 
Let us consider the regular representation of the groupoid
algebra
\[\pi_p:{\cal A}\rightarrow {\cal B}({\cal H}_p),\]
where ${\cal H}_p=L^2(\Gamma_{d(\gamma )}),\,\gamma\in\Gamma
,\,d(\gamma )=p\in E$, defined by
\[\pi_p(a)(\xi )=\frac{i}{\hbar }\xi^T\cdot M_a\]
where $\xi\in {\bf C}^n,\,n=|G|$, and the coefficient $i/\hbar $
is added to have the correspondence with quantum mechanics. To
specify $\xi$ we should remember that
\[\Gamma_{d(\gamma )}=\{(\pi_M(d(\gamma ),\lambda (d(\gamma 
)),\lambda (r(\gamma ))\}=\{(x,g_0,g):g\in G\}\] where the first
equality should be understood as the bijection.  Then $\xi
:\Gamma_{d(\gamma )}\rightarrow {\bf C}$ is given by
\[\xi (x,g_0,g)=(\xi_g)_{g\in G}.\]
\par
Let us now consider how do derivations behave under the above
representation. Let $v=v_1+v_2$ where $v_1\in {\rm O}{\rm u}
{\rm t}{\cal A}$ and $v_2\in {\rm I}{\rm n}{\rm n}{\cal A}$. If
we assume that $a\in {\cal Z}({\cal A})$ then $\pi_p(v(a))\in
\pi_p({\cal Z} ({\cal A})) \subset  {\cal Z}({\cal
B}({\cal H}_ p))$ which means that we have
\[\pi_p(v(a))=k\cdot {\bf I}.\]
\par
If $a$ is any element of ${\cal A}$, we can decompose it
\[a=a_1+a_2\]
where
\[a_1=\zeta (\langle a\rangle )\in {\cal Z}({\cal A}
).\] and
\[a_2=a-a_1 \notin {\cal Z}({\cal A}
).\]

Then
\begin{eqnarray*}
\pi_p((v_1+v_2)(a_1+a_2))\xi & = & \xi^T\cdot M_{(v_1+v_2)(
a_1+a_2)} \\ & = & \xi^T(M_{v_1a_1}+M_{v_1a_2}+M_{v_2a_2}).
\end{eqnarray*}
\par
Let us now consider the  $v_2(a_2)$-terms of the above equation
\[
\pi_p(v_2(a_2))=\frac{i}{\hbar }M_{v_2a_2}.
\]
Since $v_2 = {\rm ad}b$ for a certain $b \in {\cal A}_2 := \{b
\in {\cal A}: {\rm tr}b = 0\}$ (in such a case the choice of $b$
is unique), one has
\[
M_{v_2a_2} = M_{[b,a_2]} = [M_b, M_{a_2}],
\]
and
\[
\pi_p(v_2(a_2)) = \frac{i}{\hbar }[M_b,M_{a_2}].
\]
By taking into account that $\xi^T\bar
{X}M_{a_2}=\pi_p(v_1a_2)$, where $\bar {X}$ is an outer
derivation, we finally obtain
\begin{equation}
\pi_p((v_1+v_2)(a_1+a_2))\xi = \xi^T(f{\bf I}
+\bar {X}M_{a_2}+[M_b, M_{a_2}]).
\label{dyn}\end{equation}
By analogy with quantum mechanics we could say that if $a_2$ is
a self-adjoint element of ${\cal A}$, equation (\ref{dyn})
describes the evolution of the ``observable'' $a_2$. This
dynamical equation can be coupled with generalized Einstein
equation (\ref{Ein}) by postulating that $v$ solves equation
(\ref{Ein}) (i.e., $v_2 \in {\rm ker}{\bf G}_h$ and $v_1 \in
{\cal X}(M)$).
\par
To go from the above generalized dynamics of our model to the
usual dynamics of quantum mechanics we no longer postulate that
$v_2 \in {\rm ker}{\bf G}_h$, i.e., that equation (\ref{dyn} is
coupled to the generalized Einstein equation (\ref{Ein}), and we
assume that there exist a one-parameter family of unitary
operators $U(t) = e^{iM_bt}$. The existence of one-parameter
operator families is guaranteed by the Tomita-Takesaki theorem
but, in general, such a family depends on a state on a given
algebra.  The above postulate of the existence of $U(t)$
(independent of state) amounts to imposing on the algebra ${\cal
A}$ some further conditions (see \cite{Emergence}).
\par
Let us notice that
\[
\frac{i}{\hbar }[M_b, M_{a_2}] = \frac{d}{dt}(M_{a_2}(t))
\]
where
\[
M_{a_2}(t) = U(t)M_{a_2}U(t)^{-1},
\]
and $M_{a_2}$ satisfies the equation
\[
\frac{d}{dt}M_{a_2}(t+s)|_{t=0} = i[M_b,M_{a_2}].
\]
Since $v_1$ is any vector of $V_1$ we can choose it to be
$t$-directed; in such a case
\[
M_{v_1a_2} = \frac{\partial }{\partial t}(M_{a_2}(t)).
\]
If we assume that $M_{a_2}$ is self-adjoint and denote if by
$\hat A$, and $M_b$ is the Hamiltonian of the system and denote
it by $H$, ten the $a_2$-components of equation (\ref{dyn}) give
\[
\frac{d}{dt}\hat A = \frac{\partial }{\partial t}\hat A + [H,
\hat A]
\]
where we have assumed $\hbar = 1$. It is the Heisenberg equation
of motion well known from quantum mechanics.
\par

\section{Concluding Remarks} The model constructed in this work
is too simple to be a candidate for even a step towards the
final unification of general relativity and quantum mechanics.
However, it shows the consistency of the idea that the
noncommutative generalization of the standard geometry, when
combined with the groupoid generalization of the symmetry
concept, leads to an interesting mathematical structure having a
remarkable unifying power. Many typically relativistic and
quantum concepts smoothly cooperate with each other within this
structure (at least for a finite group $G$, and produce a
handful of valuable results. The most important of them seem to
be the following.
\par
1. Noncommutative geometry of the transformation groupoid
$\Gamma = E \times G$ is reach enough to accommodate for the
standard space-time geometry with a nontrivial contribution
coming from the group $G$ which, through its regular
representation, can be interpreted as describing the quantum
sector of our model.
\par
2. The model contributes to the understanding of the structure
of the Einstein equations. The metric is always defined on the
module of derivations, and in a more general setting these
equations are to be solved with respect to both metric and
derivations. In a usual space-time geometry, the module of
derivations is unique, and one looks for the metric. In our
model this fact is preserved in its space-time sector, but in
its quantum sector one looks for the derivations. This fact was
also signalled in one of our previous works \cite{Towards}.  It
was Madore who first demonstrated that in some derivation based
noncommutative geometries the metric could bne unique
\cite[p. 75]{Madore}.
\par
3. It is also interesting that in the quantum sector of our
model the Einstein equation has the form of the eigenvalue
equation with the cosmological constant as an eigenvalue.
\par
4. The new result is that the transition from the noncommutative
geometry of our model to the classical geometry of space-time
can be done by the averaging procedure of the elements of the
algebra ${\cal A}$. This procedure is analogous to that
typically used in quantum mechanics. The same procedure is valid
for other geometric magnitudes, such as: derivations,
differential forms, connection, metric. One can say shortly that
``averaging kills noncommutativity''.
\par
5. The transition from the dynamics of our model to the dynamics
of the usual quantum mechanics is done by restricting the model
to its quantum sector, and enforcing upon the algebra ${\cal A}$
(more strictly: upon its representation on a Hilbert space) the
existence of a one-parameter family of unitary operators.
\par
Although our model is too simple to serve as a realistic
physical model, it shows some further perspectives. It would be
interesting to explore the geometry of the dual object to the
transformation groupoid considered in the present work. If the
geometry of the groupoid is to be interpreted as giving the
``position representation'' of our model, the geometry of its
dual object could be regarded as describing its ``momentum
representation''. It seems that the natural way to construct
such a ``dual geometry'' is via making the algebra ${\cal A}$ a
Hopf algebra. This approach, by making contact with the theory
of quantum groups, and especially with the Majid program
\cite{Majid95,Majid00}, would pave the way for constructing a more 
realistic physical model.

\end{document}